%% file: paper.tex
\documentclass[runningheads]{llncs}
\usepackage[T1]{fontenc}
\usepackage{amsmath} 
\usepackage{color}
\usepackage{graphicx}
\usepackage{placeins}
\usepackage{float}
\usepackage{cleveref}
\usepackage{ulem}
\usepackage{tcolorbox}

\newcommand{\needscitation}[1]{\textcolor{red}{\textbf{\large Need Citation}}}

\begin{document}

\title{Energy Profiling of Data-Sharing Pipelines: Modeling, Estimation, and Reuse Strategies}
\titlerunning{Energy Profiling of Data-Sharing Pipelines}
%

\author{Sepideh Masoudi\inst{1} \and
Sebastian Werner\inst{1} \and
Pierluigi Plebani\inst{2} \and
Stefan Tai\inst{1}
}
\authorrunning{S. Masoudi et al.}
\institute{
\textsuperscript{1} Information Systems Engineering, Technische Universität Berlin, Berlin, Germany \\
\email{\{smi,sw,st\}@ise.tu-berlin.de}\\
\url{https://www.tu.berlin/en/ise}\\
\textsuperscript{2} Politecnico di Milano, Milan, Italy \\
\email{pierluigi.plebani@polimi.it}\\
}
\maketitle              
\begin{abstract}
\input{sections/0.abstract}

\end{abstract}
\begin{tcolorbox}[colback=white, colframe=black, boxrule=0.5mm, arc=0mm]
\paragraph*{Copyright Notice}
\textbf{Preprint}- This work has been accepted to the 37th International Conference on Advanced Information Systems Engineering(CAiSE). Copyright may be transferred without notice, after which this version may no longer be accessible.
\end{tcolorbox}
\section{Introduction}\input{sections/01_introduction}\label{introduction}
\section{Related Works}\input{sections/02_relatedworks}\label{relatedworks}
\section{Comprehensive energy profiling}\input{sections/03_energyprofiler}\label{energyProfiler}
\section{Evaluation}\input{sections/04_evaluation}\label{evaluation}
\section{Limitations}\input{sections/05.limitation}\\[1em]
\section{Conclusions}\input{sections/5.conclusion}\label{conclusion}

\begin{credits}
\subsubsection{\ackname}Funded by the European Union (TEADAL, 101070186). Views and opinions expressed are however those of the authors only and do not necessarily reflect those of the European Union. Neither the European Union nor the granting authority can be held responsible for them.
\end{credits}

%
%
%
\bibliographystyle{splncs04}
\bibliography{refs}
%

\end{document}

%% file: sections/0.abstract.tex
Data-sharing pipelines involve a series of stages that apply policy-based data transformations to enable secure and effective data exchange among organizations. 
Although numerous tools and platforms exist to manage governance and enforcement in these pipelines, energy efficiency in data exchange has received limited attention.
This paper introduces a novel method to model and estimate the energy consumption of different execution configurations in data-sharing pipelines.
Additionally, this method identifies reuse potential in shared stages across pipelines that hold the key to reducing energy in large data-sharing federations. 
We validate this method through simulation experiments, revealing promising potential for cross-organizational pipeline optimization and laying a foundation for energy-conscious execution strategies.

%% file: sections/01_introduction.tex
Companies have long recognized data as a critical business asset, and sharing these assets between organizations can unlock revenue and offer vast opportunities~\cite{Otto03072015}. However, facilitating data exchange across organizations requires both organizational and technical support.
Data-sharing pipelines represent an established method for facilitating data exchange among independent organizations~\cite{munappy2020data}. With such pipelines, source data residing within one organization is collected and transformed through sequential processing stages for utilization by a consuming organization.
These transformations are executed, leveraging pipelines as data-sharing method, according to the requirements and policies mutually established by the collaborating organizations, which may encompass privacy, analytical, or other relevant transformations.

In practice, using cross-organizational data-sharing pipelines is often hindered by technical and organizational barriers~\cite{gelhaar2021towards}, along with the lack of well-defined agreements between parties, unclear costs and resource requirements, complexity in configuration and implementation, and the need for assurances that the shared data will not be misused.
With the continued desire for open data marketplaces, several emerging governance and exchange platforms started to emerge that aim to enable organizations to manage shared data, with governance and enforcement strategies~\cite{dalmolen2020infrastructural}.
Specifically, data spaces~\cite{halevy2006principles} and federated data meshes~\cite{falconi2024data} are gaining recognition as environments that offer technical and occasional solutions. 
Moreover, the emergence of guiding principles for data providers aims to enhance the reusability of their data and the ability of machines to automatically find and use the data~\cite{wilkinson2016fair}, highlighting the complexities involved in the configuration of data-sharing.

A growing area of concern within such one-to-many (1-to-n) data-sharing pipelines is the need to assess the costs associated with data sharing~\cite{falconi2024data}. 
Particularly concerning energy expenditures, the aspiration for an energy-efficient data-sharing pipeline is paramount, as highlighted in recent studies~\cite{plebani2023teadal}.
To achieve this objective, it is essential to develop an understanding of energy consumption at each stage of the pipeline, considering factors such as computational expense, data volume, and other pertinent criteria.
Moreover, by modeling the energy consumption of stages in a data-sharing pipeline, it becomes possible to explore opportunities to reduce effort and energy consumption, as pipelines and their results can be reused when sharing the same or similar data repeatedly.
However, the possibilities and complexities in configuring these data-sharing pipelines raise the need for tools to support conscious design choices to to maximize reusability and energy savings.

Hence, we pose the following research question: \textit{How can we model energy consumption to promote energy estimation and reuse strategies in cross-organi\-zational data-sharing pipelines?}

To address this question, in this paper, we introduce a novel energy profiling model and method. We propose a comprehensive set of metrics designed to characterize energy utilization at individual stages of a 1-to-n pipeline. Each specific one-to-one (1-to-1) pipeline can be conceptualized as a configuration of steps derived from the 1-to-n pipeline model, whereby an increased number of reused steps can facilitate the evaluation and enhancement of individual and overall energy efficiency. Our model and method provide a framework for informed assessments of energy requirements within data-sharing pipelines, enabling the identification of common stages that can be shared and reused across different configurations, as well as the estimation and customization of execution configurations, which can serve to optimize energy consumption.

The remainder of this paper is organized as follows: In the next section, we review related work on energy consumption management, focusing on data-sharing pipelines and data provider services. In Section 3, we introduce the proposed energy consumption model and profiling method. In Section 4, we present a case study to evaluate the application of the model. In Section 5, we discuss the limitations of our method. Finally, in Section 6, we conclude the paper with a discussion of future work.

%% file: sections/02_relatedworks.tex
The challenges in data sharing, such as reusability, discoverablity and  programmatic actionability are among the key concerns for science and industry. 
For example, the FAIR principles~\cite{wilkinson2016fair} are one of the earliest proposals for enhancing the reusability of data holdings, with an emphasis on programmatic actionability. These principles apply not only to data in the conventional sense but also to the algorithms, tools, and workflows that led to that data~\cite{wilkinson2016fair}. 
Naturally, these principles have also inspired emerging infrastructure and platform support.
Grossman et al., for example, introduced the term 'data commons' to describe a cloud-based data platform with a governance structure that allows a community to manage, analyze, and share its data~\cite{grossman2023ten}.
Moreover, Hofman~\cite{hofman2015towards} discussed the necessity of a federated infrastructure as the solution required to construct data pipelines, allowing small and medium-sized enterprises to collaborate and share their data.
Consequently, the next challange is to increase the interoperability among different data platforms through a set of platform services and protocols for data sharing. 
Here, Hofman et al.~\cite{hofman2015towards} already hinted at the use of data sharing pipelines as one approach to archive this interoperability, however, not yet with en emphasis on energy efficiency and  reuse.

Data pipelines and their efficiency improvements are not exclusive for data sharing.
For example, Chen et al. present the real-time data processing pipeline used at Facebook. They identify five important design decisions that affect ease of use, performance, fault tolerance, scalability, and correctness\cite{chen2016realtime}. 
Additionally, Goodhope et al. discuss the design and engineering problems related to leveraging a real-time publish-subscribe system for building data pipelines\cite{goodhope2012building}.
Raman et al. explore big data pipelines as a means to break down complex analyses of large datasets into a sequence of simpler tasks. Each task features independently tuned components to enhance performance\cite{raman2013beyond}.
In these studies, the authors attempt to establish pipelines as a means to model and implement data transformation while clarifying the engineering problems and challenges in designing pipelines.

Regarding the efficiency challenges of the data management systems that pipelines utilize, Kunjir et al. present a power model based on the evaluation of different pipeline plans for processing queries~\cite{10.1145/2247596.2247648}. 
Roukh~\cite{roukh2015estimating} suggested using polynomial regression techniques for building an energy model based on pipelines. 
The authors in~\cite{ullah2022framework,10012453} investigated specific scenarios and benchmark studies to propose a non-linear relationship between energy efficiency and performance in distributed DBMS in distributed environment for optimizing the trade-offs between energy consumption and performance. 
Kurpicz et al. provided a profiling and estimation model to attribute the overall energy costs per virtual machine (VM) in heterogeneous environments~\cite{KURPICZ2018175}. This cost model considered both the dynamic energy consumption of VMs and the proportional static cost of using cloud infrastructure. 
In~\cite{GUO2024379}, the authors provided an energy consumption model for running queries and sub-queries in distributed database systems and used that model for energy consumption estimation. 
Most of the existing works also concentrate on predicting and optimizing energy consumption at the query execution level in data management systems and do not provide a comprehensive insight into energy consumption at different levels in data-sharing pipelines, such as the platform level and application level, for running data transformation pipelines.

Regarding energy profiling, Tomasoni et al. investigate the energy and network performance of different data collection frameworks (DCFs) in a mobile crowd-sensing scenario. 
They propose a methodology to profile energy consumption using an Android application, which can be used to make harvesting data from the crowd more energy-efficient~\cite{TOMASONI2018193}.
In~\cite{ZHU2019101587}, the authors suggest using profiling in commercial buildings to identify energy consumption patterns and leverage those patterns along with energy predictions for anomaly detection. Marinakis et al. propose a methodological approach for adopting big data platforms with smart energy services, gathering energy-related information from multiple sources~\cite{MARINAKIS2020572}.

Many prior studies do not consider the importance of pipeline execution and configuration in the energy consumption of the data-sharing process. Additionally, they do not investigate the potential for reusing pipelines for energy efficiency due to their multi-stage nature among different parties.

Hence, our energy profiling and estimation approach for a federated data-sharing pipeline can be seen as a more holistic method for reducing and optimizing pipelines at an organizational level, prior to the implementation of query optimization or database optimization. This contributes to the growing body of work aimed at making data-sharing systems more energy-efficient.

%% file: sections/03_energyprofiler.tex
%
%
Data-sharing pipelines are generally a set of sequential stages, each called in order during execution. 
These pipelines manage and usually transform the data between owners and consumers into a shareable and consumable form, often enacting a set of predetermined qualities and policies. 
Each Data consumer can be unique regarding its requirements, such as intervals of querying data from the owners, the level of access to the data, and data selection logic, and the rights the consumer has to that data.
Thus, sharing pipelines combine both necessary steps such as filtering, anonymization, and encryption, with quality-improving steps such as compression, formatting, and serialization into a set of stages that must be executed in the continuum between data owner and consumer to enable the sharing.

\subsection{Problem Statement}
When independent organizations with heterogeneous data policies and formats agree to share their data using pipelines, they establish a cross-organizational data-sharing pipeline, often as part of general data-sharing agreements.
In these one-to-many (1-to-n) data-sharing pipelines, data is collected and offered to specific consumer groups through a data provider. However, this data must still be transformed by a specific one-to-one (1-to-1) pipeline based on the unique requirements of each data-sharing agreement between the data consumer and the data provider.
Implementing these data-sharing pipelines should be done by the data provider organization to guarantee the correctness of transformations and transmission of data. Nevertheless, the execution of them could be done either on the data provider side or the data consumer side.

\begin{figure}
\centering
\includegraphics[width=0.90\textwidth]{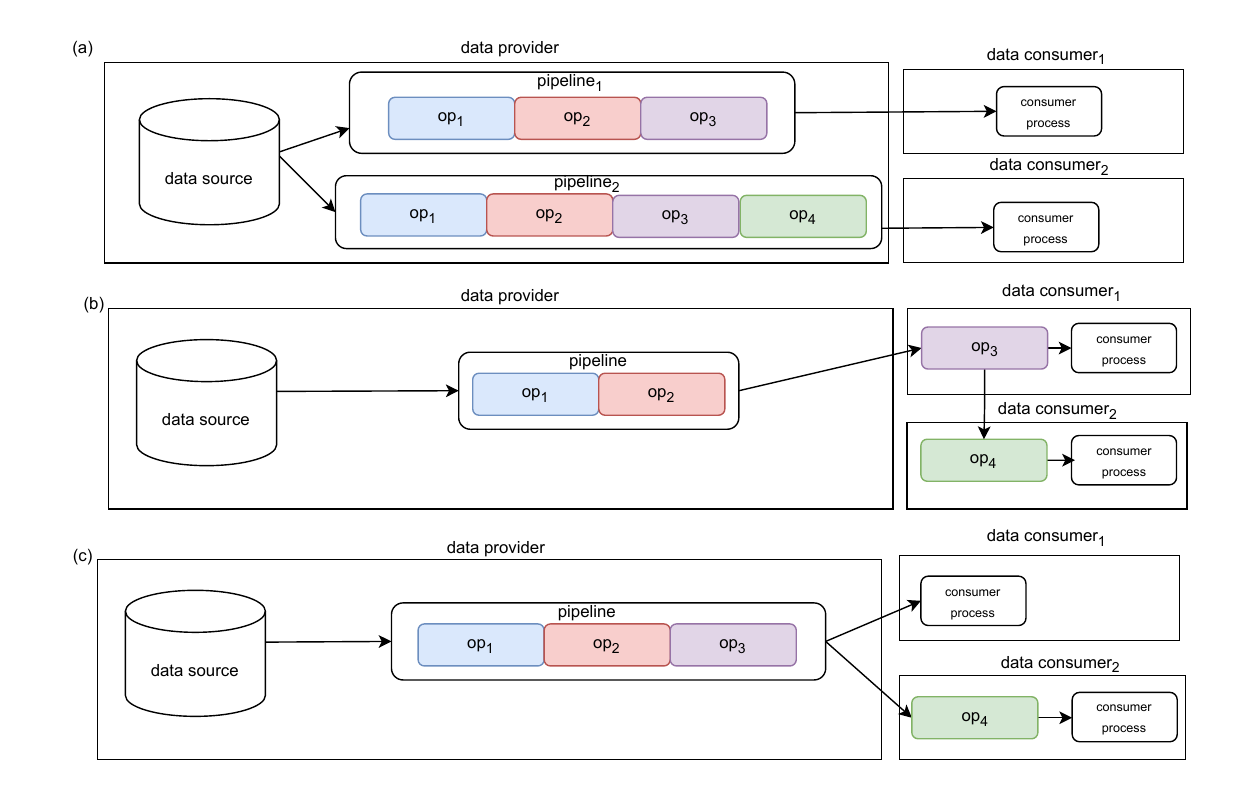}
\caption{Different execution configurations of two data-sharing pipelines.} \label{pipeConfig}
\end{figure}

Even though each one-to-one pipeline is specifically designed to meet a consumer’s requirements, some stages may be common and reusable across different one-to-one pipelines for consumers of the same type.
Hence, when designing the execution plan for a data-sharing pipeline, we need to choose from a wide range of possible configurations with different efficiencies. 
In cases where each stage of the pipeline can run on different parties' infrastructure across organizations, selecting the optimal execution configuration becomes a non-trivial decision. As shown in Figure~\ref{pipeConfig}, a data consumer (C1) can also function as a data provider for another consumer (C2) of the same type. This occurs when C1 has a higher level of access and fewer restrictions on data filtering and anonymization compared to C2, and both consumers expect similar data quality and frequency.
In these cases, a data owner can reduce and potentially save resources, cost and energy.
For example, as shown in Figure~\ref{pipeConfig}, we can create a new pipeline by reusing the common stages shared by two pipelines (op$_1$, op$_2$, op$_3$) or a subset of these stages (e.g., op$_1$, op$_2$) on the data provider side, while executing the unique stages for each consumer on the consumer side. Additionally, a data consumer can become a new data provider for other consumers. It is worth mentioning that all new pipeline configurations must adhere to governance policies and agreements between the parties involved.

However, while these data-sharing pipelines are loosely coupled and composed of independent stages, any effort to achieve energy efficiency reduction requires a comprehensive understanding of different pipeline execution configurations and the ability to estimate their energy consumption to find and propose these reuse strategies. 
To address this challenge, we investigate ways to model energy consumption that support accurate energy estimation and encourage reuse strategies in cross-organizational data-sharing pipelines.
\FloatBarrier

\subsection{Modeling the Energy of data-sharing Pipelines}\label{subsec:model}
The variety of infrastructure and hardware available to different parties in a federated data-sharing network can impact the energy consumption pipeline configurations.
Therefore, when selecting the execution configuration plan for a pipeline concerning the execution of stages across different parties and reusing common stages, we need a model that contains detailed information about the energy consumption of all executed pipeline stages at the infrastructure and hardware layers. As a first step to address this problem, we define an energy consumption model for data-sharing pipelines.

Moreover, considering the five characteristics of big data —volume, value, variety, velocity, and veracity— we assume that the volume of data impacts the energy consumption of data-sharing pipelines. This relationship is reflected in our energy model through the energy consumption associated with data transmissions at each stage.
Furthermore, data-sharing pipelines address the challenge of data variety through data transformation, while the velocity aspect is reflected in their performance.

\begin{table}
\caption{Reference table for the formula notation.\footnotesize{The units of measurement are Joules (J) for energy and Gigabytes (GB) for data volume.}}\label{deftab}
\centering
\begin{tabular}{|l|p{0.8\textwidth}|}
\hline
Abbreviation &  Definition\\
\hline
$E_{cpu}$ & Energy consumption of CPU for running the stage\\
$E_{mem}$ & Energy consumption of storage for running the stage\\
$E_{dt}$ & Energy consumption for data transmission in pipeline\\
$E_{dt\_op}$ & Energy consumption for data transmission of the stage\\
$E_{ob}$ & Energy consumption for monitoring data transmission\\
$E_{op}$ & Energy consumption of deploying and running the stage (operational energy consumption)\\
$E_{p}$ & Energy consumption of pipeline\\
$IS_{op}$  & Source input size\\
$OS_{op}$  & Output size at stage\\
$TF_{op}$  & Size transformation factor at stage\\
\hline
\end{tabular}
\end{table}

Each one-to-one pipeline can be conceptualized as a configuration of transformation stages derived from a broader set of stages used in one-to-many pipelines for the same data provider. Thus, reusing a greater number of stages can enhance both individual and overall energy efficiency.
Energy consumption in a data-sharing pipeline can be divided into three main categories. The energy consumed by deploying and running stages falls under the operational energy consumption category($E_{op}$). For operational consumption we always collect the energy consumption of all stages normalized by 1GB of input data, thus, enabling later estimation of energy consumption proportional to expected data consumption. We use $TF_{op}$ to indicate the expected mean data outcome of a stage given $IS_{op}$ for a pipeline that got 1 GB at the source.
Operational energy consumption is, thus, the sum of the energy consumption of the CPU ($E_{cpu}$) and memory ($E_{mem}$) used by each stage while running. These stages can range from complex preprocessing to simple conversions (e.g., changing the output data format). Data transmission within the multiparty environment also contributes to energy consumption. When different stages are deployed across the infrastructures of different parties, the energy consumption for data transmission may be affected due to changes in the volume of data exchanged between owners and consumers. Therefore, the total energy consumed for data transmission is one of the factors to consider ($E_{dt}$), which is a sum of the energy consumed for data transmission between every two consequential stages($E_{dt\_op}$). Monitoring the data-sharing pipeline is essential not only to ensure quality but also to record various metrics, including the energy consumption of data-sharing pipelines. Therefore, the transmission and collection of monitoring data, with a focus on energy consumption, will also incur energy consumption ($E_{ob}$). As shown in \cref{energyfmu}, the total energy consumption of a data-sharing pipeline is the sum of these three categories. Therefore, the total energy consumed by a pipeline is the sum of the energy needed for running all the stages ($\sum_{k=1}^{n} E_{op(k)}$) and the energy that would be consumed for data transmission and monitoring the data-sharing process. The abbreviations used in the formula are defined in \cref{deftab}.

\begin{equation}\label{energyfmu}
   \begin{cases}
E_{op}=E_{cpu} + E_{mem}\\
E_{dt} = \sum_{k=1}^{n} E_{dt\_op(k)}\\
E_{p}=(\sum_{k=1}^{n} E_{op(k)}) + E_{dt}+ E_{ob}\\
  \end{cases}
\end{equation}

Equation \cref{transformationfactor} defines the normalized output size \( OS_{op} \) at pipeline stage \( op \), where \( IS_{op} \) is the source input size (e.g., 1~GB) and \( TF_{op} \) is the stage-specific scaling factor. This formulation allows us to quantify how each stage transforms the data volume:  
a) Compression (\( TF_{op} < 1 \)) -- The stage reduces the data size (e.g., filtering or aggregation),   
b) Expansion (\( TF_{op} > 1 \)) -- The stage increases the data size (e.g., feature extraction or replication),   
c) Preservation (\( TF_{op} \approx 1 \)) -- The stage maintains the data size (e.g., format conversion). 

\begin{equation}\label{transformationfactor}
OS{op} = TF{op} * IS{op}\rightarrow
   \begin{cases}
TF_{op} < 1 \text{indicates compression/reduction}\\
TF_{op} > 1 \text{indicates expansion}\\
TF_{op} \approx 1 \text{indicates size preservation}\\
  \end{cases}
\end{equation}

To achieve energy efficiency in data-sharing pipelines, it is necessary to measure energy consumption at the different stages of the data-sharing life cycle. Gaining insight into the energy consumption of various stages provides the means to estimate energy consumption of different configurations and scenarios. As a result of these energy consumption estimates, configurations can be selected with energy awareness, helping to improve energy efficiency.

\subsection{Energy Profiling}\label{subsec:method}
We refer to the process of energy consumption modeling and measurement as \textit{energy profiling}, and we will explain how energy profiling can lead to improved energy consumption in data-sharing pipelines.

\begin{figure}[h]
\includegraphics[width=\textwidth]{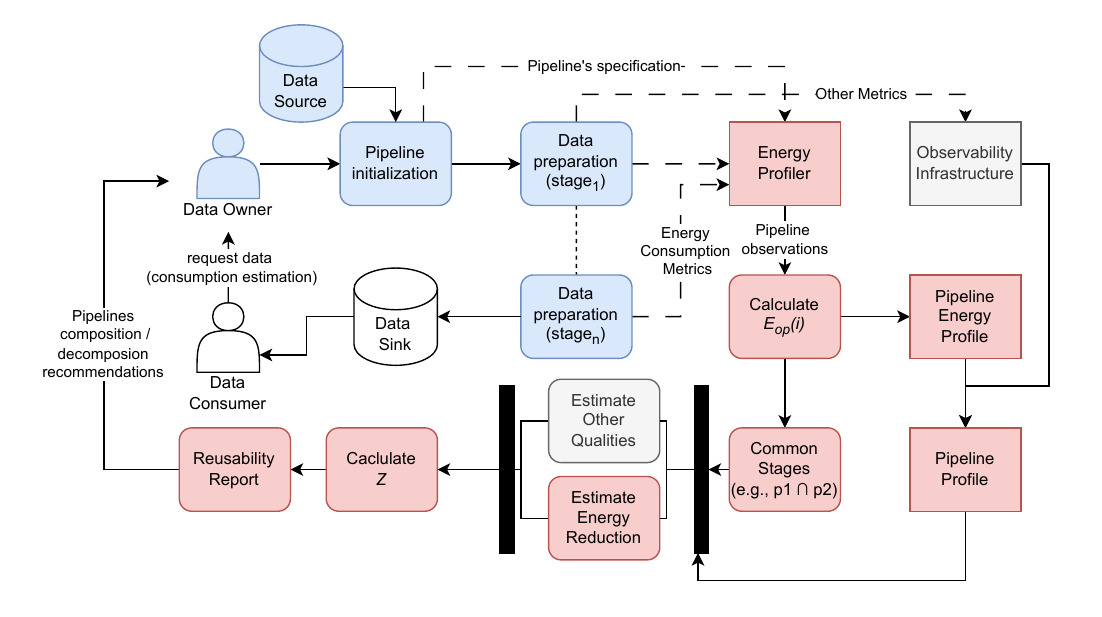}
\caption{Data flow diagram of energy profiling in data-share pipelines.}\label{dataflow}
\end{figure}

As shown in \cref{dataflow}, after the data provider defines, implements, and starts the pipeline, data preparation begins by transforming the data through multiple stages.
In this step, all stages of the pipeline are initially executed on the data provider’s side, and only the final data is forwarded to the consumer side.
The energy profiler, as a third-party between data provider and data consumer, will collect the pipeline specifications, including the sequence and type of stages that are running in the pipeline and the owner of the infrastructure on which the stages are running. 
While these stages are running, the energy profiler will collect and store energy consumption metrics for each stage to create an energy profile over time for every stage executed on the data provider’s side.
 
The data provider and data consumer administrators are responsible for deploying tools that can measure energy metrics within their infrastructure. At first, the energy profiler calculates and stores the total energy consumption of the pipeline and each stage separately using the energy model in \cref{subsec:model} and energy consumption metrics received from the data provider. Afterwards, the energy profiler will begin searching for pipelines originating from the same data provider that have common stages. 

In a data-sharing pipeline, stages that typically overlap between consumers fall into four categories: filtering, anonymization, aggregation, and converting (formatting). Among these well-known types of stages, filtering and anonymization are most likely to be common across different pipelines and reusable due to the symmetry in requirements and policies. In contrast, the other types of stages like aggregation, converting, and any other custom stages are often unique to each data consumer. Therefore, in search for common stages, the energy profiler should only consider stages that belong to the appropriate types. Subsequently, based on the energy consumption of the detected common stages, we can estimate the energy efficiency of reusing those stages in different pipelines. Thus, the final report on the energy estimation of alternative execution configuration plans for pipelines with common stages can help the system administrator optimize decisions regarding the configurations of existing pipelines to run the common stages only once. The data provider administrator can also modify the execution plan so that the unique stages of a consumer are offloaded to be executed on the consumer’s side. 

However, it is important to consider that reusing stages between pipelines and deciding where to run them can affect the data transmission volume, processing, as well as the cost and energy consumption associated with running stages in different infrastructure (e.g., running them on the consumer's side). Although these parameters are not completely predictable before trying the new configuration, as shown in \cref{commonfmu_1} and \cref{commonfmu_2}, the new configuration is most likely to be efficient if the energy consumption of the common stages is noticeably higher compared to the energy consumption of the rest of the pipeline. Moreover, by using the new configuration and continuously profiling energy consumption, the administrator of the data-sharing system can assess the efficiency of the new configuration for future use. 
Additionally, it is worth mentioning that the new execution configuration should align with the initially agreed-upon policies and rules between the data provider and the consumer. Administrators on both the provider and consumer sides can check for any possible policy violations during the negotiation phase concerning the new execution configuration and running certain consumer-specific stages on the consumer’s side.
In a real-world scenario, cross-organizational data-sharing pipelines using federated environments as infrastructure\cite{hofman2015towards,plebani2023teadal} represent a suitable underlying structure for applying the proposed method to increase reusability and to save energy.

\begin{equation}\label{commonfmu_1}
(p_{1} \cap p_{2}) \rightarrow
   \begin{cases}
p_{1} = op_{A} \rightarrow op_{B} \rightarrow op_{C}\\  
p_{2} = op_{A} \rightarrow op_{B} \rightarrow op_{D} \rightarrow op_{F}\\ 

  \end{cases}
  \Rightarrow op_{A} \rightarrow op_{B}\\
\end{equation}


\begin{equation}\label{commonfmu_2}
   \begin{cases}
E_{op(A)} + E_{op(B)} \gg E_{dt\_op(B)} + E_{op(C)} \\
E_{op(A)} + E_{op(B)} \gg E_{dt\_op(B)} + E_{op(D)} + E_{op(F)}\\ 
  \end{cases}
\end{equation}

In addition to energy, other factors such as performance, response time, and the volume of transmitted data may influence the selection of a new execution configuration. While the primary focus of our method is on energy efficiency, it is essential to incorporate a mechanism that allows the administrator to evaluate the feasibility of reusing a subset of stages in the new execution configuration or altering their execution location. This evaluation should be based on insights provided by other methods or monitoring tools, such as the performance ~\cite{hassan2025managing}, data quality ~\cite{smithintegrating}, deployment planning of stages~\cite{amado2025automated}, the average response time of a stage, and the pipeline's sensitivity to delays.
As illustrated in \cref{feasibilityfmu}, the administrator can assign an impact weight to each factor, enabling the consideration of other qualities (e.g., performance) alongside energy in the decision-making process. This means that, as shown in \cref{dataflow}, if reusing common stages is expected to have a significantly negative impact on other qualities—such that it outweighs the estimated energy savings—the reuse of those stages will be excluded.
In \cref{feasibilityfmu}, \textbf{W} represents the impact weight assigned to energy and other qualities, while \textbf{Q} denotes the estimated value of metrics other than energy. The value of \textbf{Q} can be positive (indicating an expected improvement) or negative (indicating an expected degradation). If the result of \cref{feasibilityfmu} is negative, the administrator may decide to exclude the reuse of related common stages due to constraints imposed by other qualities.

\begin{equation}\label{feasibilityfmu}
\begin{cases}
W > 0 , E_{r} > 0 , Q \in {R}\\
\\
Z= (W_{E}*E_{r})+\sum_{k=1}^{n} (W_{k}*Q_{k})\\
\\
Z>0 \rightarrow \text{consider reusing common stages}\\
Z<0 \rightarrow \text{skip}\\
\end{cases}
\end{equation}

\FloatBarrier

%% file: sections/04_evaluation.tex
For the evaluation of the proposed method regarding energy model and reuse strategies, we implemented a preliminary simulation of federated data-sharing pipelines. The goal of this running example is to examine the feasibility of the presented energy model and demonstrate the impact of energy profiling on improving energy awareness and estimating energy consumption for different pipeline execution plans. To the best of our knowledge, our approach is the first to propose a systematic way to measure and estimate energy consumption in data-sharing pipelines, making comparisons with other approaches impossible at this time.
An energy profile for each pipeline consists of the total energy consumption of each stage of that pipeline on its current infrastructure. Moreover, the energy profile also demonstrates a list of common stages shared between that specific pipeline and other pipelines (including the identifiers of those pipelines), along with the percentage that shows the portion of common stages' energy consumption relative to the overall energy consumption of the pipeline.

The validation is being performed on a synthetic dataset of pipeline stages stored as a \texttt{*.csv} file. The stages fall into four categories: filtering, anonymization, aggregation, and converting. Each row in the dataset (each stage) contains information about the energy consumption of that stage at different levels, as defined in this study (shown in Table~\ref{operationtab}). The amount of energy consumption per unit (e.g., CPU unit, data transmission unit, etc.) is specified as static metrics at the beginning of the experiments. Therefore, the total energy consumption is estimated by multiplying the resource usage amount by the energy consumption of that resource. This simulation is open source and available on GitHub\footnote{GitHub address: https://github.com/Sepide-Masoudi/Energy-profiling-in-federated-data-sharing-pipelines/tree/master}. The implementation and execution of this model are not restricted to the use of a specific software. In a real-world scenario, these metrics can be measured using well-known monitoring and observation tools. 
Kepler\footnote{https://sustainable-computing.io/} is a well-known energy consumption exporter that exposes statistics from applications running in Kubernetes clusters at the container and node levels. The metrics measured by Kepler cover a wide range, including CPU, GPU, and storage energy consumption\cite{Keplersite}. 
Also, Tapo P115\footnote{https://www.tp-link.com/de/home-networking/smart-plug/tapo-p115/} is a smart monitoring socket. 
Moreover, the tool introduced in~\cite{masoudi2025pre} can also be utilized to implement the method proposed in this paper, demonstrating its feasibility for potential real-world experiments in the future.

\begin{table}
\caption{Information stored for each stages in dataset.}\label{operationtab}
\begin{tabular}{|l|l|}
\hline
Name & Definition\\
\hline
stages\_id & stages's unique identifier\\
stages\_type & stages's type(filtering, anonymization, aggregation, converting)\\
cpu\_usage\_unit & cpu consumption for running the stage\\
memory\_usage\_unit & storage consumption for running the stage\\
output\_data\_volume & the volume of output data of the stage\\
input\_data\_volume & the volume of input data of the stage\\
observation\_unit & resource usage for monitoring the stage\\
\hline
\end{tabular}
\end{table}

\begin{figure}
\includegraphics[width=\textwidth]{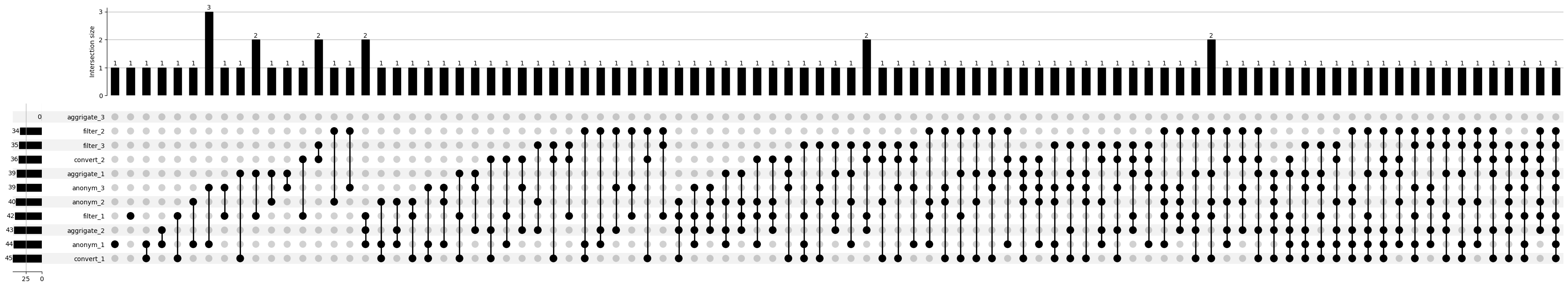}
\caption{Aggregation of stages across different pipelines in a single simulation round, with 100 pipeline populations created from random combinations of 11 stages from 4 types(filter, anonymization, aggregation, convert).} \label{pipelineOperatoinPlot}
\end{figure}

The running example simulates energy profiling, and the dataset of stages (.csv file) could be the output of any energy metrics collector. 
In this validation, the capabilities are implemented using Java. All experiments are performed on a MacBook Pro M3 with 36 GB of RAM and an Apple M3 Pro chip. The creation of data-sharing pipelines is automated, with pipelines being randomly populated based on the stages in the dataset (.csv file). The total number of stages in each pipeline, and which stages should be executed, are selected randomly. In the next step, for every pair of pipelines, common stages in the same order are identified, and the energy consumption of these common stages is calculated using the energy model explained in \cref{subsec:model}. In~\cref{pipelineOperatoinPlot}, the plot shows the common stages between different pipelines in one iteration of the validation.

In the final report of the validation, as shown in~\cref{operationtab}, the administrator can view a list of pipelines with common stages between them, the energy consumption of those common stages, the total energy consumption of each pipeline in its current configuration, and the percentage showing the portion of energy consumption attributed to common stages relative to all stages in the pipeline. This report helps the data-sharing system administrator gain a comprehensive understanding of potential alternative execution configurations that could be shared among multiple data consumers, as well as the amount of energy that could be saved by creating a new pipeline for the common stags and reusing them across different consumers. Therefore, in a real-world scenario where the number of pipelines and operational complexities are greater, leveraging energy profiling enables the system administrator to identify opportunities for improving energy efficiency by running common stages only once and estimating the percentage of energy consumption for the shared components relative to the total.
This estimation takes into account various factors, such as the specific resource usage of each stage on the current infrastructure, the energy associated with data transmission between stages, and the energy consumption from monitoring those stages (shown in~\cref{operationtab}).

\begin{table}
\caption{Part of energy profiler report about the common stages and energy consumption of them in one iteration of the validation ($E_{CO}$: common stages total energy consumption,$E_{P_{A,B}}$: total energy consumption of pipeline).}\label{operationtab}
\centering
\begin{tabular}{|l|l|l|l|l|l|}
\hline
$P_{A}$ & $P_{B}$ & $CO$ & $E_{CO}$ & $E_{CO}/E_{P_{A}}$ & $E_{CO}/E_{P_{B}}$\\
\hline
3,4,5	& 2,3,4 & 1,5 & 2911 & 30\% & 84\%\\
2,3,5,6,9,10 & 2,3,5,6,7,9 & 2,3,5,6 & 27542 & 13\% & 24\% \\
4,5,6,8,10 & 3,4,5 & 4,5 & 8638 & 1,7\% &91,5\% \\
\hline
\end{tabular}
\end{table}

Through this experiment, we confirmed the feasibility and benefits of using the proposed energy model and energy profiling in identifying and highlighting reuse opportunities. The model successfully calculated the energy consumption of a pipeline, broken down by its stages, detected common stages across different pipelines, and estimated the potential energy savings from reusing those stages. This demonstration validates that the model can enable data provider and data consumer administrators to make informed decisions on optimizing execution configurations, thereby minimizing overall energy consumption through the reuse of common stages or offloading specific stages to the consumer's side.

%% file: sections/05.limitation.tex
The proposed model represents an early-stage effort to reduce waste in data-sharing environments. Consequently, it relies on several assumptions that require validation in real-world settings. Nonetheless, it is anticipated that ongoing developments in academia and industry -- particularly in data mesh architectures~\cite{datamesh}, data spaces~\cite{halevy2006principles}, and federated data sharing~\cite{falconi2024data} -- are progressing toward sufficient maturity to enable such validation in the near future.

The primary objective of this work was to model the reuse and energy-saving potential in data-sharing pipelines. To assess the feasibility of the approach, pipelines were modeled as sequences of fine-grained, easily distinguishable operations, inspired by prior work~\cite{munappy2020data}. While this abstraction enables the intended optimization, it also limits the immediate applicability of the model, as current pipeline frameworks typically do not expose such granular and clearly defined operation boundaries.

Furthermore, the model assumes that the energy consumption associated with data operations and transfers can be profiled precisely. In practice, such measurements are subject to environmental variability~\cite{SmartWatts}, interference from co-located processes (e.g., noisy neighbors), and instrumentation limitations, which often result in only approximate values. Accordingly, future extensions should incorporate probabilistic representations of energy consumption to more accurately account for such variances.

Finally, while the simulation conducted in this work demonstrates the potential of the proposed method, its generalization remains to be validated. On the one hand, federated data-sharing environments are still in early stages of adoption, and the supporting pipelines are not yet sufficiently mature for full-scale application. On the other hand, existing platforms do not currently accommodate both the requirements of federated data sharing and the specific capabilities assumed in the presented model. However, it is expected that this gap will diminish as such platforms evolve. For example, the TEADAL approach~\cite{plebani2023teadal} envisions a compatible environment and is currently undergoing evaluation. Application of the proposed method within such a setting is planned for future work.

%% file: sections/5.conclusion.tex
In this study, we propose a method for modeling the energy consumption of data-sharing pipelines when independent parties share their data. Additionally, we introduce a method to improve the energy efficiency of data-sharing pipeline execution through energy profiling and the reuse of common stages between different pipelines. We evaluate our approach by running an example of energy profiling on synthetic pipelines and demonstrate how this method can assist with energy estimation in various configurations and utilized by other tools for energy saving.

In future studies, we aim to extend the energy profiling method by using machine learning algorithms for the estimation process, enabling more accurate predictions of total energy consumption for pipelines in different configurations and reuse strategies. This will allow us to account for changes in parameters such as the infrastructure on which the pipelines run and provide estimations for elements that are difficult to measure, such as energy consumption related to data transmission over the network. We also plan to consider the cost of different configurations, in addition to energy consumption, in future studies. Finally, we will apply the energy profiling method in a real-world data-sharing context to investigate the impact of the proposed approach in a practical use case by developing a tool that leverages this method to optimize pipeline execution planning.